\documentclass[11pt,a4paper]{article}
\usepackage{graphicx,psfrag}
\usepackage{color}
\usepackage{bm}
\usepackage[utf8]{inputenc}
\usepackage[T1]{fontenc}
\usepackage{soul}
\usepackage[thinspace,thickqspace,textstyle]{SIunits}
\usepackage{ctable}
\usepackage{dcolumn}
\usepackage{multirow}
\usepackage{placeins}

\newcommand{\beqn}{\begin{eqnarray}}
\newcommand{\eeqn}{\end{eqnarray}}

\newcolumntype{d}{D{.}{.}{-1}}

\newcommand{\e}[1]{\cdot 10^{#1}}

\title{Transfer conditions and transmission bias in capillaries of vacuum interfaces}
\author{
Laurent Bernier\footnote{TU Berlin, Germany}, 
Markus Taesch\footnote{TU Berlin, Germany}, 
Stephan Rauschenbach\footnote{University of Oxford, Dept. of Chemisty},
Julius Reiss\footnote{TU Berlin, Germany}}
\begin{document}

\maketitle


\section*{Abstract}

A detailed study of the transfer of ions in transfer capillaries of electrospray ion sources is presented. 
The laminar flow field for various capillary sizes and wall temperatures is calculated. 
It forms the base of ion transfer simulation of a large number of ions with space charge. 
This allows to study the thermodynamical conditions of the ions during transfer, which are found to vary strongly with the capillary dimensions. 
The dependence of mass flow and ion current on the size of the capillary is presented.
Simple scaling relations are derived and tested. 
The method also allows to predict a transfer bias between different ion species depending on the difference in ion mobility and the composition of transferred ions.  
\\[1em]
keywords: electrospray ion source, transfer capillary, detailed simulation, atmospheric interface, transfer bias, transfer conditions, thermal stress   \\[1em]


\newpage

\section{Introduction}

Electrospray ionization (ESI) is a universal, charged particle source with manifold applications in bio-chemical and physical analytical research, generating molecular ions over the entire range of small organic molecules to folded proteins and even entire viruses.~\cite{Konermann_AC85(2013),Snijder2013}

The properties of the generated ion beam, i.e., quantity and condition of the species of interest, are decisive for the application. 
Indiscriminate ionisation and transmission is key to quantifying results in many analytical applications, whereas native mass spectrometry~\cite{Benjamin1998} is dependent on gentle, well-controlled conditions in the source region, which guarantee the retention of the solution conformation in the gas phase.

In ESI source, molecules undergo ionisation, release from a desolving droplet, transport into vacuum, moving along with an increasingly diluted gas flow through transfer capillaries, pinholes, skimmers and ion optics. 
The presence of external electric fields, space charge fields, and complex gas flow, which includes strongly varying velocity as well as temperature and pressure, creates a complicated and convoluted system, which cannot be described holistically at present.

Therefore, in recent years, key elements have been investigated for relevant properties.
The first step after the ESI, the vacuum transfer, being the transport of the ions from ambient pressure to a much lower pressure environment through a capillary or pinhole, is of major importance as its efficiency imposes an upper limit of the total efficiency of the experiment.
Often high losses occur in this initial part of the apparatus so that an improvement here offers an equivalently large overall gain.
In addition, as the ions are transported from ambient condition to vacuum, the environmental parameters change by orders of magnitude, which can seriously affect the properties of the ions (charge state, conformation, velocity) through temperature gradients, collisions or fields.

The gas flow in the capillary, caused by the pressure gradient and significantly influenced by the heating, dominates the transport of analyte ions.
During its transfer, an ion experiences complex thermodynamical conditions (pressure and temperature) which influence its mobility.
Potentially, the interactions with the gas flow can be responsible for modification of the ions such as unfolding, declustering or fragmentations during the transfer due to heating and intense ion-neutral collisions.
It is, therefore, the initial step of the study to describe the gas dynamics and hence the thermodynamic conditions in detail.

One of the first major studies  of the ion transport through long and narrow capillaries itself  has been published by Lin and Sunner \cite{lin1994ion}, already comparing the measurements with analytical estimates and numerical models.
It attributes the behavior to laminar gas flow driven transport, with losses due to space charge and diffusion. 
The characteristics of the capillary (length, diameter, overall shape) as well as its heating were found to affect the behavior of ESI significantly.

More detailed experimental investigations through current- and mass spectrometry measurements of the capillary and the electrospray interfaces reveal the dependencies on various parameters like spray position, capillary heating, intensity, pressure differential, and properties  and capillary temperature. 
These are presented in \cite{page2007ionization} by  measuring currents and assigning the losses to different parts of the interface. Significant improvements through optimization of the interface could be obtained.~\cite{pauly2014hydrodynamically} 
Not least the transport of different ion species was investigated and a bias in the transmission was found in~\cite{page2009biases}, depending even on the presence of the other ions, hence suggesting an interaction between transferred ions.

Using a one dimensional description of a simple gas flow through a capillary with an  empirical turbulence model, we could show in a previous publication~\cite{bernier2018gas}
that the transition between laminar and turbulent flow seems possible in the typical transport regime, making it a very good candidate to explain the wide range of transmittance from few \% 
\cite{page2009biases} to unity \cite{pauly2014hydrodynamically,KrutchinskyPadovanCohenChait2015}. 
Quasi 3D ion transport simulation showed the interplay of space charge, diffusion and gas flow and allowed us to rationalize how it affects the transmittance.~\cite{bernier2018gas}

The successful modeling however showed limitations of this approach where many key influences were introduced as semi-empirical parameters.
In particular, 3D details of the gas flow as well as the heat flux through the walls are crucial to the transport behavior.
The dependence of the heat flux on the experimentally accessible wall temperature could only be estimated.  
High transmissions are found for laminar flow, with a good quantitative agreement.
Since high transmission rates are technically most relevant, in the following we focus on laminar flows.

Here we simulate the spatially resolved flow and thermodynamical variables to validate and improve the results of the previous study.~\cite{bernier2018gas} 
We confirm these findings and investigate aspects unattainable with the previous approach. 
Namely, the friction dominated gas flow, especially important for low heating, and the strongly 
stratified temperature distribution were not covered. 
This allows to refine the understanding of the gas flow, and to simulate on this base transmission rates and transmitted currents. 
Strong gradients in the flow towards  the wall, not only in the velocity but also in temperature and density, can be quantified, which have important consequences for the condition of transport of the ions.
Furthermore we present our results for various capillary lengths and diameters typically used in experiments to elucidate the impacts of those dimensions on the practically relevant properties.
Finally, in this context we discuss the interactions of simultaneously transported ion species and show how the transmission of these mixtures is affected and find biases effects similar to those experimentally reported in~\cite{page2009biases}.\\

Different teams also developed tools to model and simulate the behavior of the ions during their transport through the inlet capillary.
A simulation of  a rough vacuum chamber with  a skimmer includes a simulation  of a heated capillary without ions, however choosing an unrealistic slip-wall boundary condition   \cite{gimelshein2014numerical}. 
In two recent, connected contributions the gas flow   \cite{skoblin2017gas}  and the ion transport in a capillary \cite{skoblin2017numerical} are investigated.
Laminar flow is assumed and a good agreement  with experiments is found.
Especially, it was shown that due to space charge effects, a maximal transmitted current exists, agreeing with our findings \cite{bernier2018gas}.
In contrast, in  \cite{wissdorf2016gas} a comparison of flow rate and  elaborate optical  measurements with effective gas flow models lead to the conclusion  that a turbulent flow seems to exist.
The length of the capillary in both cases differs (\unit{6}{\centi\meter} versus about
\unit{18}{\centi\meter}) whereas similar
diameters are considered, which is probably, at least partly, responsible for the differences exposed.
This shows that the nature of the flow strongly depends on details of the considered  system. As stated
in~\cite{wissdorf2016gas}, turbulence might not be the only relevant factor to predict the overall
transmission of the ions, but it strongly influences the behavior of the ions inside the
capillary inlet, given that the flow conditions and thus the consecutive entrainment of the particles
is different.

This article investigates the capillary utilized in many setups to transfer ions from  atmospheric conditions to the first reduced-pressure part of the machine.


\section{Methods}

The simulations of the ion transport are performed in two steps: 
First, the gas flow is computed and second, the produced gas flow field is used to simulate the transport of the particles through the transfer capillaries. 
This separation exploits the fact that the particles have a marginal influence on the underlying flow.~\cite{bernier2018gas}

\subsection{Gas Flow Simulation}
\label{sec:method_flow}

Capillaries of various length and diameters are investigated. 
The general geometry as well as the coordinate system used in the following are represented in Fig.~\ref{fig:temperature}-a.
Non-slip walls conditions with either a prescribed wall temperature or an adiabatic wall (no heat flux) are considered.
At the inlet, a funnel of parabolic cross section and of \unit{1.0}{\centi\meter} length is added, identical to the geometry used in the experiment described in \cite{pauly2014hydrodynamically,bernier2018gas}.
At the start of the funnel, the flow velocities are so small that the conditions are well approximated by ambient parameters, using a so called non-reflecting boundary condition.~\cite{BaylissTurkel1982}

Similarly, another parabolic funnel (which is not present in the experiment) is added at the end of the capillary.
It is used to mimic the influence of the first pumping stage by a pressure forcing term, using a so called \emph{sponge} area. 
We tested the results to be insensitive to the enforced pressure due to the supersonic nature of the capillary flow, which disrupts the transfer of information upstream (see supplementary material for details).

Several geometries were simulated for comparison, varying in the diameter of the different parts as well as in the length of the capillary (a complete list is provided in sup. mat.)

The gas flow simulations are performed using a compressible Navier-Stokes solver with a skew-symmetric formulation and implementation as presented in~\cite{reiss2014conservative}.
The domain is fully axial symmetric and laminar flow is assumed by which the flow itself becomes axial symmetric. 
This allows to use modified axial symmetric equations  reducing the problem effectively to two dimensions. 
The simulated gas is air, with the assumption that it follows the ideal gas law and that the other gaseous substances such as the vaporized solvent do not significantly modify the properties, as stated in~\cite{bernier2018gas}.\\

\subsection{Particles Simulation}

The ions are modelled as individual particles.
This was found to be superior to a particle density model.~\cite{WissdorfPohlerKleeMullerBenter2012}

The equation of motion is applied to each individual particle.
 It models the entrainment by the underlying gas flow, the Coulomb forces between the charged particles
and the molecular diffusion:
\begin{equation}
\mathbf{x}_p^{n+1} = \mathbf{x}_p^{n} + 
\Delta t \left(\mathbf{u}_f^{n} + K^{n} \mathbf{E}^{n}\right)
+ \Delta \mathbf{x}_\mathrm{diff}\,
\end{equation}
Here $\mathbf{x}_p^n$ is the current position of the particle,  $\Delta t$ the considered time step,
$\mathbf{u}_f$ the gas velocity, $K$ the ion mobility, $\mathbf{E}$ the computed electric field
and $\Delta \mathbf{x}_\mathrm{diff}$ the displacement resulting from a random walk modeling the diffusion.
The upper indexes $n$ and $n+1$ indicate two consecutive time steps for the position of the particles or any other parameter evaluated at the corresponding position.

This simple dynamics assumes that the acceleration phase of the ions is short in comparison with the considered time scale. 
The particle response time, describing a typical time span of the acceleration phase,
is several orders of magnitude lower than the  typical convection time (\unit{\power{10}{-10}}{\second} vs. \unit{\power{10}{-4}}{\second}).
The entrainment of the particles by the gas flow imparted by $\mathbf{u}_f^{n}$ dominates the transport and is, thus, of essential importance.
Hence, the velocity of the ions is given by the asymptotic value described by the fluid velocity plus a drift velocity induced by the local electric field.

This electric field effect is taken into consideration by computing the Coulomb forces between the simulated particles.
The computation of the electrical forces is performed using the Fast Multipole Method (FMM), described in~\cite{greengard1997new} and implemented in the ScaFaCoS library \cite{scafacos-web}.
The computation of the electrical field assumes that no relevant magnetic
field is generated for the low currents and charge velocities considered.

In order to further reduce the computational requirements, an artificial charge factor is used for the computation of the emitted electrical field to simulate the
presence of several ions, represented by only one particle in the simulation.
The electrostatic boundaries of the system are not accounted for. 
As a consequence, the constant potential on the walls of the metal capillaries is not enforced, large variations of the field along the capillary are not expected.

The ion mobility $K$, representing the proportionality between electric forces and the particle drift velocity, depends on the nature of the buffer gas and on the thermodynamic conditions (pressure $p$ and temperature $T$) and is given by
\begin{eqnarray}
K = K_0 \frac{T}{T_0} \frac{p_0}{p}\,. 
\end{eqnarray}
Macroscopic ion diffusion is modeled via a random walk producing Brownian motion.
After each time step $\Delta t$, the position of each particle is shifted by
\begin{equation}
\Delta l = \left(2 \Delta t \frac{k K T}{q}\right)^{\frac{1}{2}}\,,
\end{equation}
in random direction,  equal in the three Cartesian space dimensions because of the
isotropy of the diffusion process.
It depends on the time step and on the diffusion constant, which relies on the thermodynamical state of the gas via the ion mobility.
Further, $k$ is the Boltzmann constant, $T$ the temperature, $q$ the electrical charge of the particle.

The ions are inserted into the simulation at the start of the straight capillary ($z=0$). 
Two different distributions are considered for the location of the new particles: a flat one, where the ion density is constant over the capillary section, and a Gaussian one, leading to a normal distribution centered on the capillary axis. 
In both cases, the inserted ions are stretched along the axial direction for the transport length of one time step, in order to avoid the creation of ion packages for each time step.  
For the Gaussian distribution, the standard deviation is based on the radial dimension of the tube:
$\sigma = 5\cdot10^{-2}\, R$.
The required particle flux is given by the desired electrical current, the individual charge of the ion species considered and the artificial
charge factor. 

Particles are considered to be transmitted when they reach the right end of the axial direction,
\emph{i.e.} when they leave the capillary alongside the gas. 
They are considered as lost when they leave the domain at any other boundary.
Especially, this implies that the ions are considered as lost as soon as they hit the walls of the capillary.
While justified in the case of metal capillaries, this is certainly a simplification of the physico-chemical processes occurring in general, as for instance described in~\cite{wissdorf2016gas}.


\section{Results and Discussions}
  
\subsection{Gas Flow}
\label{sec:result_gasflow}

The gas flow in the capillary, caused by the pressure gradient and significantly influenced by the heating, dominates the transport of analyte ions.
During its transfer, an ion experiences complex thermodynamical conditions (pressure and temperature) which influence its mobility.
This allows potential modifications of the ions, heating and molecular collisions leading 
to unfolding, declustering or fragmentation. Therefore, the first step towards a better 
understanding of such processes relies in the detailed description of the gas dynamics and the 
resulting thermodynamic conditions.

\begin{figure}
	\begin{center}
		\includegraphics[width=\linewidth,clip=true]{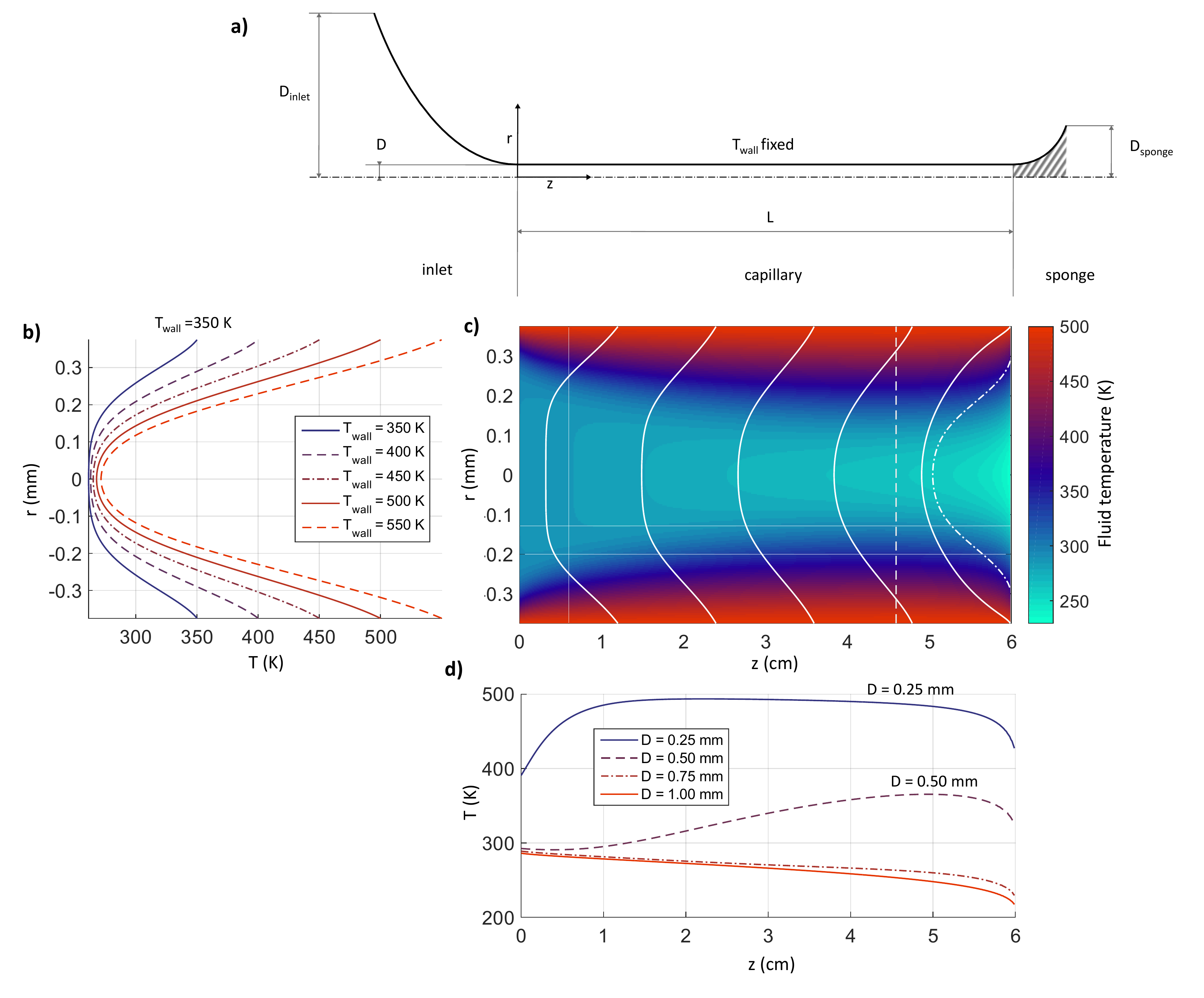}
	\end{center}
	\caption{The temperature: a) Schematic representation of the simulated domain. The coordinate system and the different length defined here are used throughout this article. b) Radial temperature profiles. The data is taken from the simulation of \unit{6.0}{\centi\meter} long capillaries with a diameter of \unit{0.75}{\milli\meter} and plotted at $z \approx  \unit{4.5}{\centi\meter}$ (dashed line in figure c). c) Temperature field within a \unit{6.0}{\centi\meter} long and \unit{0.75}{\milli\meter} wide capillary. Dimensions are re-scaled. The wall temperature is set to \unit{500}{\kelvin}. The solid lines represent the temperature profile at the positions where they touch the wall. The dash-dotted line is the sonic line, indicating the limit between subsonic flow (left) and supersonic flow (right).  d) Axial temperature profiles plotted at the capillaries axes. All domains are \unit{6.0}{\centi\meter} long, to reveal the influence of the diameter.}
	\label{fig:temperature}
\end{figure}

\begin{figure}
	\begin{center}
		\includegraphics[width=\linewidth,clip=true]{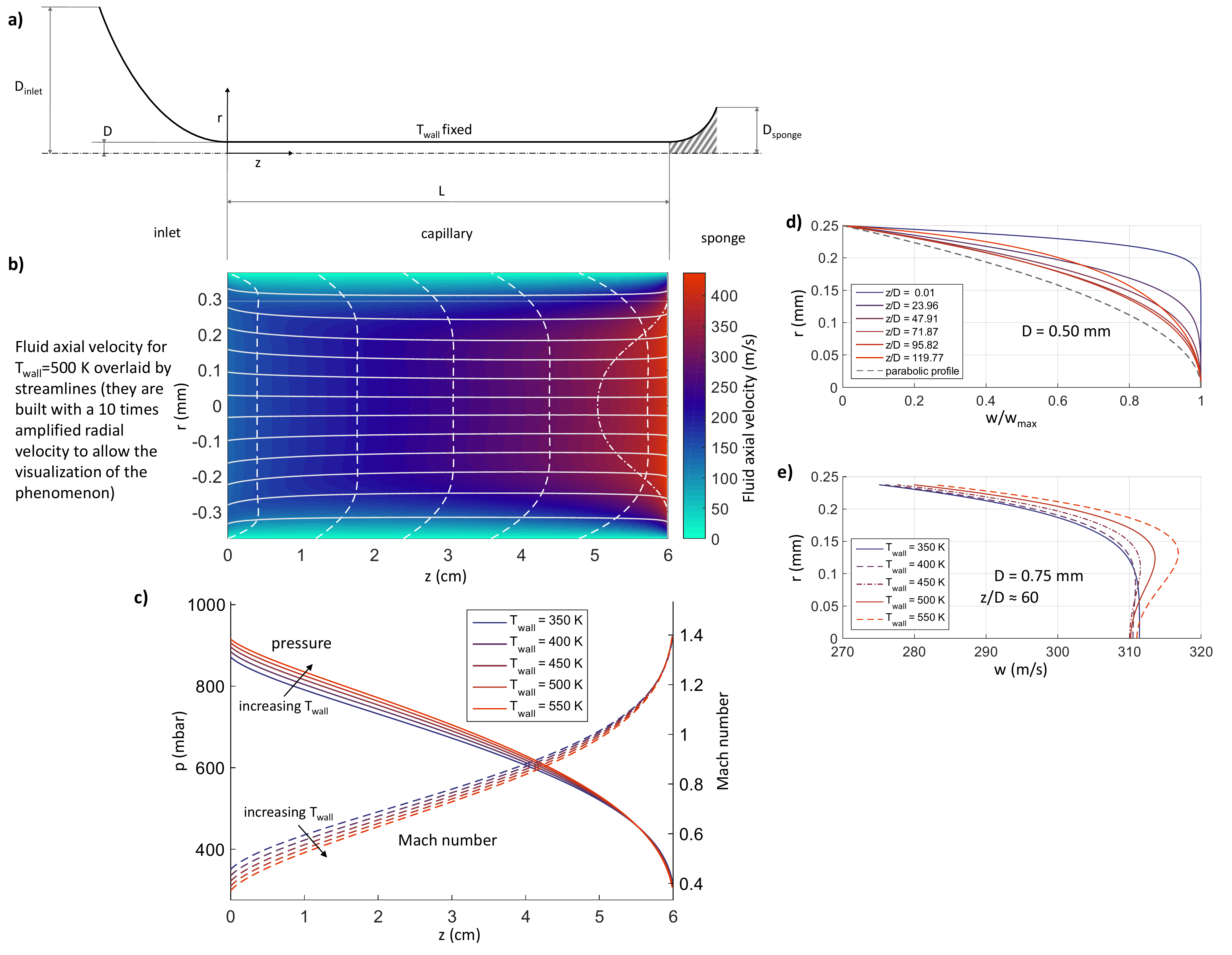}
	\end{center}
	\caption{Velocity and pressure: a) Schematic representation of the simulated domain. b) Axial velocity field within a \unit{6.0}{\centi\meter} long and \unit{0.75}{\milli\meter} wide capillary. The wall temperature is set to \unit{500}{\kelvin}. The dashed lines represent the axial velocity profile at the positions where they touch the wall. The dash-dotted line bending inwards on the right side is the sonic line. The solid lines represent the streamlines of the flow, where the radial effects have been artificially amplified to make it visible. Dimensions are re-scaled. c) Pressure (straight lines) and Mach number (dashed lines) axial profiles at the symmetry axis, showing the influence of the wall temperature for a \unit{6.0}{\centi\meter} long and \unit{0.75}{\milli\meter} wide capillary. d) Radial profiles of the axial velocity component, normalized with the maximal velocity reached. The capillary considered is \unit{6.0}{\centi\meter} long with a diameter of \unit{0.50}{\milli\meter}. Adiabatic non-slip wall conditions are considered and the profiles show the evolution along the capillary in comparison with a parabolic profile (dashed line). e) Radial profiles of the axial velocity around the axis depending on the wall temperature. The capillaries considered are \unit{6.0}{\centi\meter} long with a diameter of \unit{0.75}{\milli\meter}. The profiles are plotted at an axial position of $z \approx \unit{4.5}{\centi\meter}$}
	\label{fig:velocity}
\end{figure}

A typical heated flow produced by the simulation is presented in Fig.~\ref{fig:temperature}-c and~\ref{fig:velocity}-b.
Temperature and velocity fields show strong stratification towards the axis.
This strong change of state variables close to the wall is referred to as \emph{boundary layer}, a consequence of the no-slip boundary conditions for the velocity and the prescribed temperature at the wall.

\subsubsection*{Temperature field} 
Figure~\ref{fig:temperature}-b presents radial temperature profiles of the gas in a capillary (length $L=\unit{6.00}{\centi\meter}$ and diameter $D=\unit{0.75}{\milli\meter}$) for different wall temperatures, taken at $z\approx \unit{4.5}{\centi\meter}$, i.e. after the gas had traveled most of the capillary length.
A significant difference between the temperatures of the wall and the center of the capillary is observed.
Over the entire capillary length the cooling effect due to acceleration of the gas compensates the heating from the wall (see Fig.~\ref{fig:temperature}-b).
The extent of this phenomenon depends on the diameter as illustrated in Figure~\ref{fig:temperature}-d.
Capillaries of large diameter show a continuous drop in temperature along the axis ($r=0$), whereas for the narrow $D=\unit{0.50}{\milli\meter}$ pipe, a slight increase of the temperature is observed. 
For the narrowest pipe considered, the temperature at the center reaches values close to the wall temperature shortly after the inlet.
This can be explained by the fact that the temperature boundary layer thickness does not strongly depend on the radius of the pipe. (See Supp. Mat.) 
Thus, when the radius is smaller than the typical boundary layer thickness, the heat flux from the wall is sufficient to increase the temperature at the center. 
Otherwise the cooling accompanying the pressure drop dominates. 
With values of the capillary diameter as  typical used in MS instrumentation yielding different heating regimes, seemingly small changes in instrumentation  can have strongly different flow conditions.

\subsubsection*{Choking and sonic conditions}

The pressure difference between the ambient conditions and the first pumping stage is driving the gas flow and hence the ion transport.
Because of its great magnitude, we expect the flow velocity $u$ to accelerate, reaching the speed of sound $c$ at the end of the capillary. 
Classical gasdynamics theory demands a minimum in cross section to accelerat beyond the speed of sound. This assumption can be used as a boundary condition to define the gas flow in the tube, as done in our previous work.~\cite{bernier2018gas}.
However, this picture is not entirely correct. 
In fact, the sonic line, where the gas velocity equals the speed of sound ($\mathrm{Ma} = u/c =1$), is located well inside the capillary, by about 13 diameters in the case here discussed. 
Similar effects are discussed in~\cite{murphy1984effects}. 
This effect is understood as a consequence of the varying thickness of the boundary layer. Because of the strong pressure gradient, the streamlines form a (converging-diverging) nozzle-shaped pattern nearing the capillary exit. This builds up an effective, so called, Laval nozzle for the flow and the standard behavior is observed: the sonic condition is reached where the effective diameter is minimal, after which the gas is accelerated to supersonic velocities.

Figure~\ref{fig:velocity}-c presents the Mach number along the capillary axis for a diameter of \unit{0.75}{\milli\meter} and the different temperatures imposed at the wall.
In all cases, the Mach number reaches a value of about 1.4 at the end of the capillary. 
The pressure curves are depicted in Fig.~\ref{fig:velocity}-c, agreeing well with the measurements presented in \cite[p. 247]{PrandtlOswatitschWieghardt1993}.
The axial pressure profiles clearly characterize the phenomenon called chocking. 
The pressure reached at the end of the capillary does not correspond to the pressure in the connected chamber (usually around 1 to 10~mbar) but is found to be approximately 300~mbar in all cases.
Chocking means that only a finite quantity of gas can be transfered through a pipe,  defined by the area at sonic conditions. 
Once this condition is reached, further lowering the pressure at the outlet
does not change the flow behavior. 
Given the very high pressure difference in the considered case, this blocking regime is always reached.

This has two noteworthy consequences. 
First, the capillary flow is independent of the pressure used in
the first stage, which simplifies the investigation of the capillary.
Second, the amount of gas flowing into the machine is independent of the pumping pressure, for pressures below the value of approximately 30\% of the ambient value in this case.
This is an important aspect in understanding the behavior of a capillary atmospheric pressure interface: the flow behavior in the capillary cannot be influenced by changing of the pumping.
Chocking also implies that the flow rate only weakly depends on the length of the capillary.

\subsubsection*{Velocity field}

The quantity linking ion and gas motion is the velocity field of the gas flow, which is influenced by heating as well.

First looking at the behavior without heating, Fig.~\ref{fig:velocity}-d presents the normalized axial velocity profiles at different positions along the capillary ($D=\unit{0.50}{\milli\meter}$ and $L=\unit{6.00}{\centi\meter}$). 
At the beginning of the capillary, the profile presents a plateau shape and gets closer to the classical parabolic shape further downstream, behavior typical for the initial stage of capillary flow.~\cite{PrandtlOswatitschWieghardt1993} 
In all simulated cases (different diameters and length), one observes a similar process leading towards a parabolic velocity profile across the tube. 
A small deviation at the outlet is the consequence of the decreasing boundary layer and near-sonic flow.
A certain capillary length is necessary  for the parabolic profile to develop. 
This length is 200 to 300 measured in diameters according to the empiric formula given in~\cite{PrandtlOswatitschWieghardt1993}, which is not fulfilled  for all investigated capillaries.

Figure~\ref{fig:velocity}-e presents radial profiles of the axial velocity of heated capillaries for different wall temperatures at $z\approx\unit{4.5}{\centi\meter}$.
For high wall temperatures, the maximal velocity is found off the axis.

 While this effect is small and no fundamental consequences for the ion transport are expected, it is surprising and thus worth a brief discussion.
Strong heating produces high temperatures close to the wall and, thereby, low gas densities.
By this, the same pressure gradient along the capillary accelerates the lighter gas close to the wall to a greater extent than the heavier gas in the center, explaining the altered velocity profile.

\subsection{Particles Transport Simulations}

The gas flow simulations are the base for simulating the ion transport through the capillary.
Assuming a low density of ions compared with the gas molecules, the feedback of the ion motion on the gas is very small and is further neglected.~\cite{bernier2018gas} 
Thus, ion transport simulations with varying ion parameters can be performed for each flow configuration.
As a result of an ion transport simulation, we can investigate the absolute transmitted currents or local transmission rate. 
In particular, simulations allow the elucidation of the influence of design parameters and operation conditions, as well as the ion properties and the mixture of different ion species.
Furthermore, it is possible to extract the conditions of every modelled ion individually during its transfer (e.g. experienced temperature, surrounding pressure).

In the following sections, the transport of one molecular species, Rhodamin~B ($m=443, z=1$) is simulated.
When the influence of a mixture  of species is considered, we used Bradykinin~$3^+$ ($m=1062.9, z=3$),
Neurotensin~$3^+$ ($m=1675.5, z=3$) and Fibrinopeptide~A~$2^+$ ($m=2307.0, z=2$) ion, which all present  m/z of  the same order, yet  have different charge state and different mobility, as well as a very well described dataset, which is available to be compared against our predictions.~\cite{page2009biases}
To account for different inlet geometry, i.e. funnel shaped or conventional flat interface, two different ion distributions are investigated.
A cloud of ions close to the axis of the capillary, modelled with a Gaussian distribution, 
reproduces the collimated ion cloud injected via a funnel shaped inlet with immersed ESI-emitter as described in \cite{bernier2018gas}. A flat distribution represents conventional setups where the capillary samples ions from an electrospray located in front of the straight capillary entrance.

The movement of the ions in axial direction is dominated  by the gas flow velocity which is much larger than the drift velocity induced by the electric field. 
By contrast, in radial direction, the gas velocity is near zero (as can be seen from the stream lines in Fig.~\ref{fig:velocity}) so that the space-charge-induced drift velocity dominates, visible as the expansion in  Figure.~\ref{fig:particleHistory}-a..
Together with diffusion, which plays a role in proximity to the wall where ion concentrations gradients are high, this leads to an expansion of the ion cloud during the transfer.
The expansion rate increases with space charge, which is mainly determined by the incoming ion current.

\subsubsection{Thermal Conditions During Transfer}
\label{sec:result_particles}
\begin{figure}
	\begin{center}
		\includegraphics[width=\linewidth,clip=true]{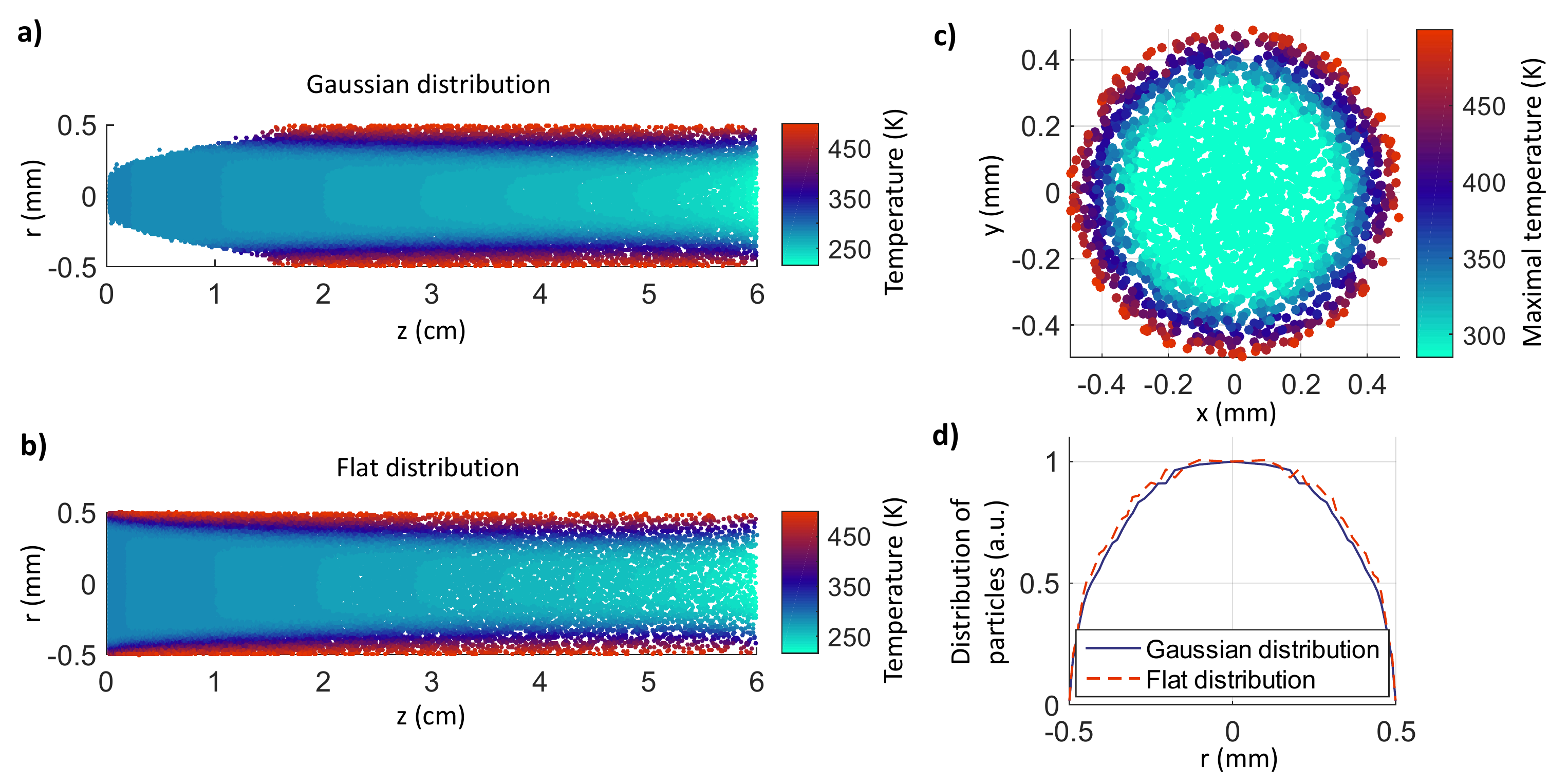}
	\end{center}
	\caption{a) and b) Temperature at the position of the simulated ions during transfer for a Gaussian (a) and a flat (b) inlet distribution. c) Position of the transmitted particles in the outlet area colored with the maximal temperature encountered by the ions during their flight. A Gaussian initial distribution has been used. d) Radial distribution of the transmitted particles at the outlet for a Gaussian and a flat distribution of the ions set in the inlet area normalized to one at the axis. In all cases the capillary is \unit{6.00}{\centi\meter} long and \unit{1.00}{\milli\meter} wide and the temperature at the walls set to \unit{500}{\kelvin}.}
	\label{fig:particleHistory}
\end{figure}

The simulation of a large number of individual particles allows describing transfers of individual ions through the capillary and extracting data related to their environment during the transfer, as well as net transmission characteristics.
The results presented in the following represent the simulation of the larger diameter (\unit{1.00}{\milli\meter}) and the medium length (\unit{6.0}{\centi\meter}) tubes.
The wall temperature is set to \unit{500}{\kelvin}. Snapshots of both simulations with positions of the ions and local gas temperature are presented in Fig.~\ref{fig:particleHistory}-a and b.\\
In figure~\ref{fig:particleHistory}-d, the radial distribution of the transmitted particles is represented, considering the two different ion distributions in the inlet, as discussed before: Gaussian and flat. 
The simulations presented had been performed with the same incoming current of
\unit{60}{\nano\ampere}, leading to different transmitted currents: \unit{8.90}{\nano\ampere} for
the flat distribution \emph{vs.} \unit{26.88}{\nano\ampere} for the Gaussian one.

While  a stark contrast in the transmitted current can be found, no major qualitative difference can be seen for the shape of the normalized distributions.
In both cases, the entire section of the capillary contains particles and the
density of particles near the axis is almost constant. 
Close to the walls in contrast, a steep decrease of the ion density is observed, due to the ion loss process occurring at the wall.

The temperature of the gas surrounding the transported ions is a parameter of great relevance, because it has an influence on the chemical and/or conformational state. 
In particular for native mass spectrometry~\cite{Benesch_ChemRev(2007)}, the retention of the folded state critically depends on the conditions surrounding the ion during transfer. 

In a heated capillary interface, the thermal profile strongly depends on the  geometry.
The flow presents a boundary layer in the gas temperature field in the radial direction. 
Due to the relatively short length of the capillary, the temperature in the middle of the tube continuously decreases along the axis despite the applied heating.
This is visible in Fig.~\ref{fig:particleHistory}-c, which displays the maximal temperature surrounding the ions during their transport through the capillary.
For the large proportion of the particles, located in the central area, the maximal temperature corresponds to the (ambient) temperature, i.e. the temperature at the beginning of the capillary.
This implies that those particles never reached a position close enough to the walls to feel the influence of the heating\footnote{Heat radiation, which is  not considered here, might have an effect on the ions.}.
Further, the clear reproduction of the shape of the boundary shows that the probability to be transferred back closer to the capillary axis by diffusion is very low for particles reaching the area of higher temperature, as the space charge effects only contribute to displacements towards the walls and the background gas velocity has a negligible radial component.
This is consistent with the fact that diffusion plays a less important role in the overall transfer process, unless for particles in the region close to the walls or when the space charge effects are small (\emph{e.g.} for very low incoming currents).

The  mixture of ion species  available in the target experiment will be further affected by the ion transport in the subsequent stages of the apparatus, especially the first pumping stage.
The outer, most heated ions are expected to be in the outer part of the  jet expansion taking place at the outlet of the capillary.
A skimmer located on the same axis as the capillary is thus likely to remove the heated ions.
In contrast, an ion funnel might be able to guide the outer and hence most heated ions to the next pumping stage.

\subsubsection{Influence of the Geometry}
\begin{figure}
	\begin{center}
		\includegraphics[width=.75\linewidth,clip=true]{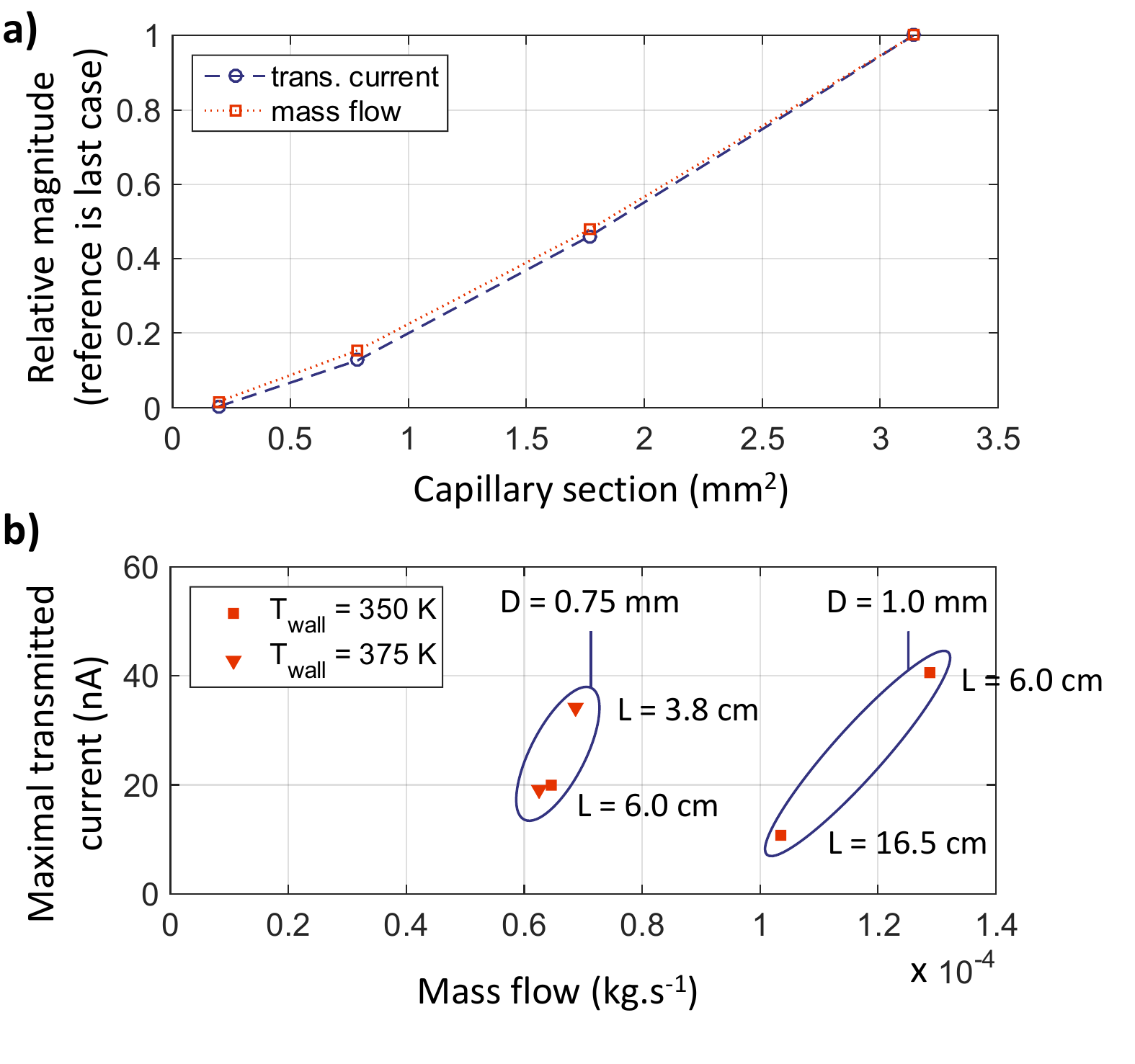}
	\end{center}
	\caption{a) Evolution of the transmitted current and mass flow for different capillary diameters and hence cross sections, with a wall temperature of \unit{500}{\kelvin} and a length of \unit{6.00}{\centi\meter}. The reference values are taken for the largest capillary (corresponding to $D = \unit{1.0}{\milli\meter}$) b) Evolution of the transmitted current for different capillary lengths. Results for two capillary diameters and two lengths each are presented, with respect to the mass flow. The first case presents results for a \unit{0.75}{\milli\meter} wide capillary and a wall temperature of \unit{375}{\kelvin}. The second a \unit{1.0}{\milli\meter} wide capillary and a wall temperature of \unit{350}{\kelvin}.}
	\label{fig:transmittedCurrent_geom}
\end{figure}

In the picture developed so far, the strength of each described factor strongly depends on the geometry of the capillary.
For understanding a given or specifying a new  vacuum interface, both the maximal ion current and the gas load are defining parameters.
Figure~\ref{fig:transmittedCurrent_geom}-a presents the influence of the diameter. 
It shows that the change of transmitted current and mass flow is similar with this parameter, both scaling with the cross section ($A = \frac{\pi}{4} D^{2}$).
The boundary layer effects become dominant for the thinnest capillary (diameter of \unit{0.25}{\milli\meter}), reducing gas flow and ion transport likewise. This also explains the deviation from a linear relation through zero, as the boundary layer reduces the effective cross section, emphasizing the importance of the chocking.\\
The standard Hagen-Poiseuille flow predicts a proportionality between flow rate and $D^4$, and not with $D^2$ as found here, which is explained by the large pressure drop and the resulting limitation of the velocity by chocking.
This linearity can be explained with basic considerations. The mass flow is estimated as the product of the cross section and the average flow velocity $u_{\mathrm{av}}$.
At the inlet of the capillary, the Mach number close to the axis is roughly 0.5, so the velocity is half the ambient sound velocity. The boundary layer being still thin here, this can be used as the average value over the cross section.
The maximal current is estimated based on the space charge drift (see supplementary material), yielding 
\begin{equation}
    D^2 \sim 
    \frac{I_{\mathrm{max}} }{u_{\mathrm{av}}}
    \frac{ L              } {u_{\mathrm{av}}}.
\end{equation}
Here, $I_{\mathrm{max}}$ is the maximal current, where the first factor determines the influence of space charge and the second one the transport time.

Not only the diameter but also the length of the capillary $L$ plays a significant role. Figure~\ref{fig:transmittedCurrent_geom}-b shows the evolution of mass flow and ion current for varying lengths. Two sets with different diameters are presented.
In both cases, the change in ion current is much larger than the change of mass flow rate. 
For example, in the \unit{1.0}{\milli\meter} diameter case, 
the mass flow decreases by a factor of about 1.25 while the current decreases by a factor of four.

The small change in flow rate is again explained by the chocking of the flow, which determines 
the flow rate by the sound velocity at the chocking point and effective cross section, suggesting no influence of the capillary length.  
The mild change visible is due to boundary layer growing slowly with the length and the larger total heat flux through  the larger wall surface. Decreasing the wall temperature for a long capillary to roughly match the heat flux of a short tube ($T=\unit{375}{\kelvin}$, $L=\unit{3.8}{\centi\meter}$) in Figure~\ref{fig:transmittedCurrent_geom}-b shows the influence of this effect.

The strongly reduced ion current is explained to the most part by the longer time for the ion cloud to expand.
The slightly decreased flow velocity adds to this. These values agree very well with the above estimate. For the capillary with $D=\unit{0.75}{\milli\meter}$, the current is inverse proportional to the length. For the \unit{1.0}{\milli\meter} capillary, the change of the flow rate has to be considered in addition.

The effects described are independent of the incoming ion distribution. 
If an ion cloud with a flat distribution expands, the ions far away from the axis will reach the wall and be lost, the ions close to the axis do not expand sufficiently during transfer and are thereby transmitted.
These two parts can be considered as separated by the ions reaching the wall just at the end of the capillary. The electric field induced by the outer ions does not influence the inner ions due to the symmetry of the distribution. 
This means that the inner ions behave just as a sharp distribution of matching initial density. 
Even if such a case is always possible in practice, the ion distribution cannot have an influence on the maximal current transmitted. However, this shows that for a given transmitted current, the total initial current (and hence the loss rate) is much higher for a flat distribution.  

\subsubsection{Influence of the Heating and Current}
\begin{figure}
	\begin{center}
		\includegraphics[width=.75\linewidth,clip=true]{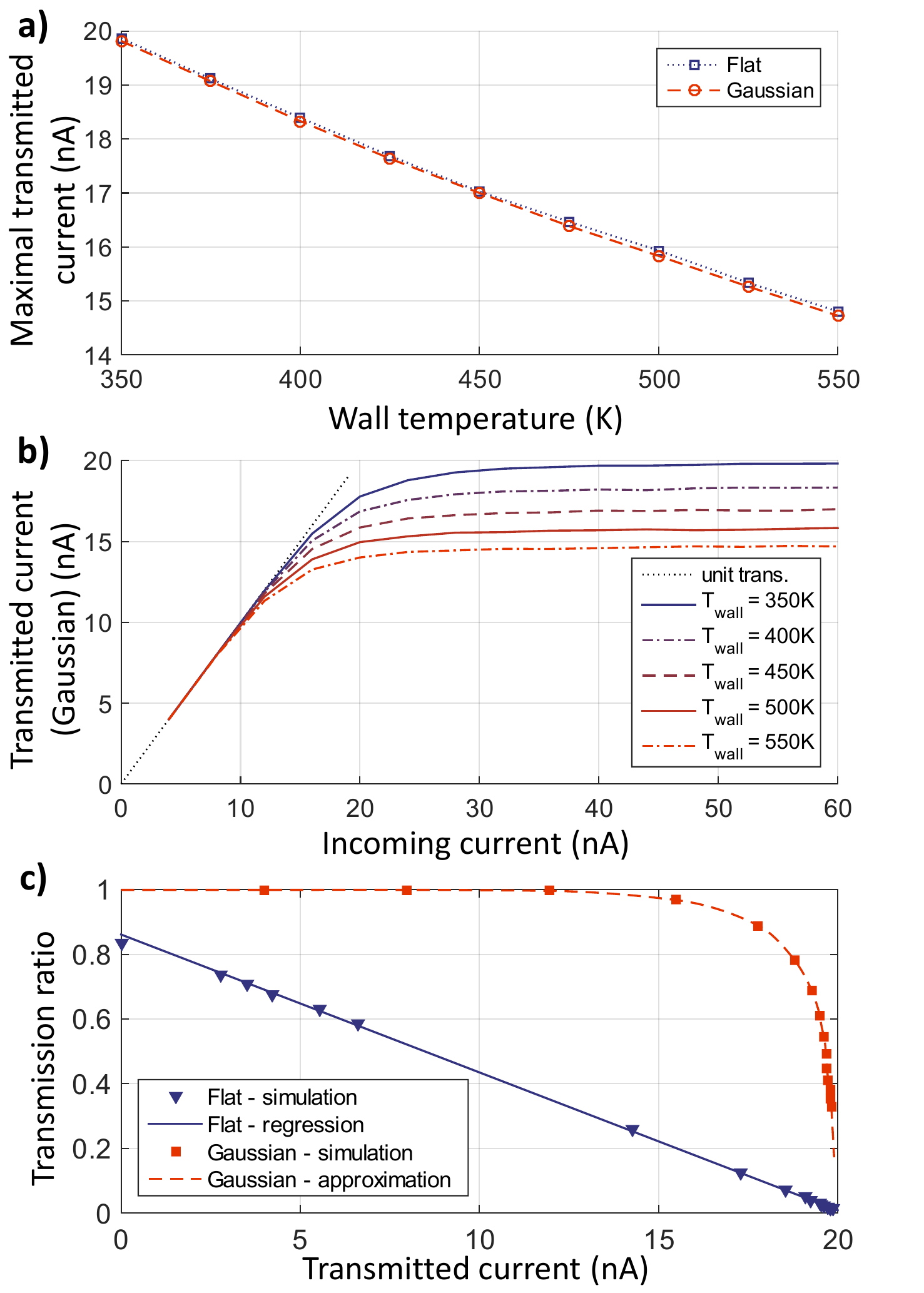}
	\end{center}
	\caption{All here presented results are computed with a capillary length of \unit{6.00}{\centi\meter} and a diameter of \unit{0.75}{\milli\meter}. a) Maximal transmitted current of incoming current of \unit{60}{\nano\ampere} (Gaussian distribution) and \unit{1512.5}{\nano\ampere} (flat distribution) b) Transmitted current for an initial Gaussian distribution depending on the wall temperature, in comparison with the unitary transmission (dashed line) c) Transmission ratio with respect to the transmitted current for a Gaussian and a flat distribution. Wall temperature is \unit{350}{\kelvin} in all cases. }
	\label{fig:transmittedCurrent}
\end{figure}
The dependence of the transmitted ion current on the heating of the capillary is shown in Fig.~\ref{fig:transmittedCurrent}-a 
The relation found agrees qualitatively with previous work,~\cite{bernier2018gas}.
The maximal, transmitted currents, obtained for both initial distributions, are identical. 
Nevertheless the transmission depends on the heating, yielding a slight decrease for increasing  wall temperatures. 
As the gas velocity decreases when the enforced wall temperature decreases, the ions have more time to expand to the wall. This effect is further supported by higher effective ion mobility close to the walls due to higher temperature. 
However, the maximal transmitted current is achieved for dramatically different incoming currents: \unit{60}{\nano\ampere} for the Gaussian distribution ($\sigma = 5\cdot10^{-2}\, R$) and \unit{1500}{\nano\ampere} for the flat distribution.
This means that sharping the ion distribution at the inlet can lead to a much higher transmission efficiency, up to unitary transmission for low incoming currents (Fig.~\ref{fig:transmittedCurrent}-b).
The saturation observed for high incoming currents at all temperatures, also observed experimentally e.g. in~\cite{bernier2018gas}, is a direct consequence of the space charge effects.
Additional charge introduced at the inlet beyond the saturation limit only leads to higher losses.\\
Following the estimation of the transmitted current presented in the supplementary material, and assuming a flat distribution of ions over the inlet cross section, we can approximate the transmission ratio as:
\begin{equation}
    \delta = 1 - I_\mathrm{trans} / I_\mathrm{max} 
\end{equation}
with $\delta$ the transmission ratio and $I_\mathrm{trans}$ the transmitted current. 
The maximal transmitted current is 
\begin{equation}
I_\mathrm{max} =     \frac{\pi \varepsilon_0 u_{\mathrm{av}}^2 R^2}{2 K  L},
\end{equation}
with 
$K$ the  mobility of the considered species,  $\varepsilon_0$ the vacuum permittivity, $u_{\mathrm{av}}$ an averaged axial velocity of the gas flow and $R$ the radius of the capillary. A similar estimate is derived in \cite{skoblin2017numerical}. 
It shows a linear dependence of the transmission ratio with respect to the transmitted current. The simulations presented in Fig.~\ref{fig:transmittedCurrent}-c reproduce this behavior with good accuracy. 
The comparison with corresponding results for a Gaussian distribution shows that this kind of measurement characterizes the ion distribution. 

Two aspects are, however, not taken into account in the estimation. First, diffusion has not been considered, which however plays a significant role for low charge densities and hence low transmitted currents. 
Second, the estimate assumes equal charge and current density in the inlet, and by this a constant gas velocity. However, the boundary layer in the gas velocity field  induces a very small  ion current  close to the capillary walls. This leads to some accumulation of ions in the simulation in this region and increased losses. This effect explains why the regression line, and thus the transmission ratio, does not reach the value 1 even for low currents. A more realistic distribution would involve proportionality between local charge density and flow velocity. In practice, this effect is  small as the boundary layer in the inlet region is  very thin, and is expected even thinner in the  case of conventional capillaries without funnel shaped inlet, so that this deviation should be small in experimental realizations. 
   
\subsubsection{Transport of ion mixture}
\begin{figure}
	\begin{center}
		\includegraphics[width=\linewidth,clip=true]{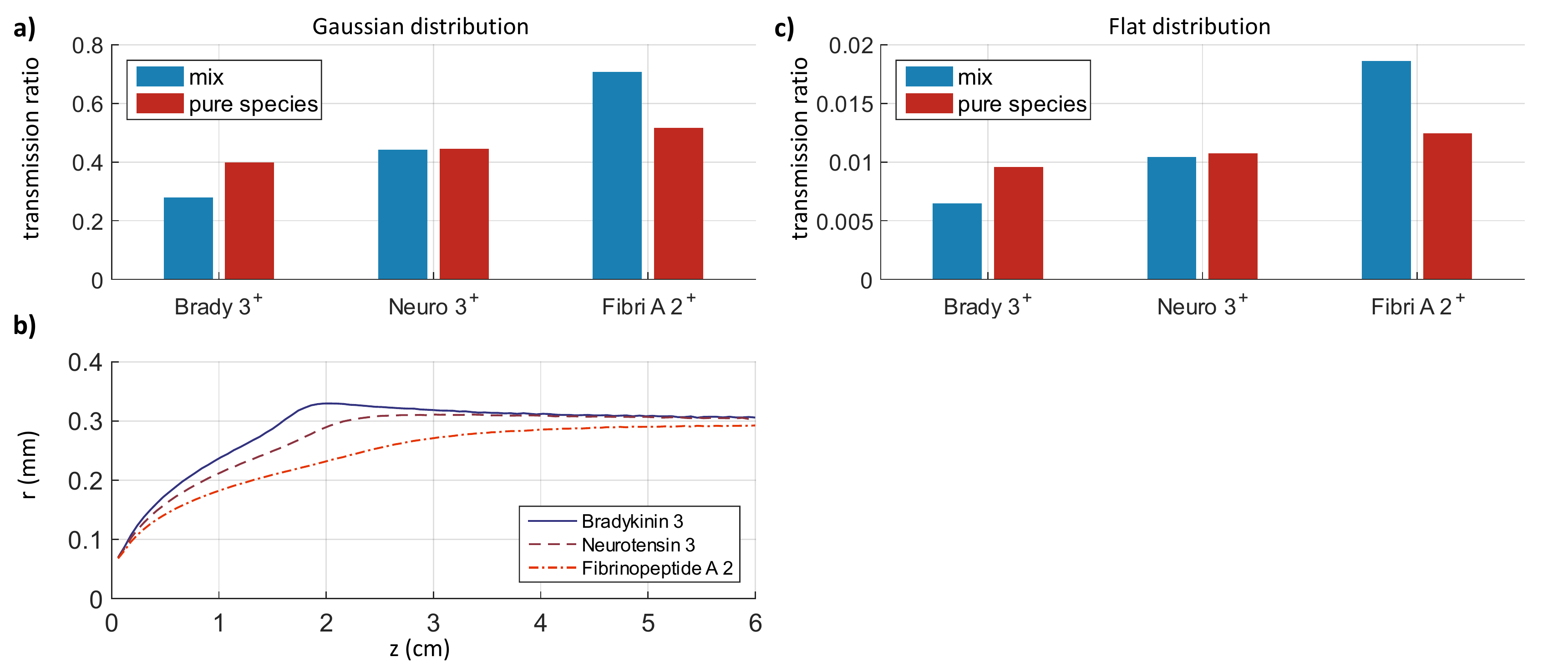}
	\end{center}
	\caption{Representation of the bias effect. Comparison of the transmission ratio depending on the species, in each case in the mix and for the species alone. a) Ions are introduced with a Gaussian distribution. b) Comparison of the mean radius of the ions along the capillary for the Gaussian inlet distribution. c) Ions are introduced with a flat distribution. All simulations are performed for a \unit{6.0}{\centi\meter} long and \unit{1.00}{\milli\meter} wide capillary and an incoming current of \unit{60}{\nano\ampere} for the Gaussian distribution and \unit{2.5}{\micro\ampere} for the flat distribution.}
	\label{fig:transmissionBias}
\end{figure}

\begin{table}	\centering
	\begin{tabular}{@{}ccccc@{}}
		\toprule
		Species                & $F_\mathrm{e}$ for Gaussian & $F_\mathrm{e}$ for flat & Reduced mobility                & m/z    \\
		                       & distribution                & distribution            & (cm$^2\,$ V$^{-1}\,$  s$^{-1}$) &        \\
		\midrule
		bradykinin $3^+$       & 2.018                       & 1.908                   & 1.352                           & 354.3   \\
		neurotensin $3^+$      & 1.840                       & 1.810                   & 1.207                           & 558.5   \\
		fibrinopeptide A $2^+$ & 1.602                       & 1.776                   & 1.033                           & 769.0   \\
		\bottomrule
	\end{tabular}
	
	\caption{Enhancement factors computed as the ratio of the transmitted current in the short capillary
		(\unit{3.8}{\centi\meter} long) towards the long capillary (\unit{6.0}{\centi\meter}). The 
		diameter is \unit{0.75}{\milli\meter} in all cases. The incoming current is 
		\unit{60}{\nano\ampere} for the Gaussian distribution and \unit{2.5}{\micro\ampere} for the flat 
		one. The reduced mobility is for a N$_2$ buffer gas but used here in air for qualitative results.
	}
	\label{tab:res_enhancementFactor}		
\end{table}

In a realistic scenario, several species of ions are mixed in the beam in contrast to the simulations we used so far to describe the general behaviour with a single species involved in the transmission. Each species is characterized by an individual mass, mobility and charge state, which determine their transport behaviour. In addition, the charge leads to mutual interactions between the different fractions of the beam which can cause a shift in transmission relative to the behaviour of the pure species.
This effect is called transmission bias \cite{page2009biases}, i.e. the transmission is specific for each species involved, which is further dependent on the parameters of the ion source such as length and diameter of the capillary or gas flow rate.
Based on the experimental results presented in \cite{page2009biases}, here we employ the simulation methodology that has been developed in this work to quantitatively explain transmission bias and its dependency on geometry and parameters of the ion source.\\
We compute transmission bias as the ratio of the transmitted currents at the outlet of the capillary for the different lengths considered, which yields an enhancement factor $F_\mathrm{e}$ for each species: 
$F_\mathrm{e} = \frac{I_\mathrm{trans,long}}{I_\mathrm{trans,short}}\,,$ where the transmitted current $I_\mathrm{trans}$ for each species is considered for the long as well as for the short capillary with the same emitted current.
The results are presented in table~\ref{tab:res_enhancementFactor}. Both inlet distributions deliver similar results, consistent with the observations presented in~\cite{page2009biases}. 
The enhancement factor is higher for the species with higher reduced ion mobility. 
The exact contribution of the different processes possibly involved in the establishment  of this effect is complex to define. 
For the example of bradikinin~$3^+$, neurotensin~$3^+$ and fibrinopeptide~A~$2^+$, simulations have been performed and the results are presented in Fig.~\ref{fig:transmissionBias}-a and c. 
In the case of a Gaussian inlet, the radial distribution found in the simulations for each ion species in the tube (Fig.~\ref{fig:transmissionBias}-b) indicates that high mobility species will be transported towards the wall more quickly, thereby rapidly reducing the space charge that species of lower mobility experience, keeping them closer to the axis. 
When charge saturation is reached, this phenomenon leads, for both inlet distributions, to a lowering of the transmission ratio of bradykinin~$3^+$ when considered in a mix rather than alone and to an increase of the ratio for fibrinopeptide A $2^+$.
However, transmission bias does not affect the charge saturation value of the system. 
In consequence, it cannot be used to increase the absolute quantity of a species considered in a mix where it has the smallest mobility, compared to the same species alone.
Further, since increasing the length of the capillary has a stronger negative impact on the transmission of the species with a high reduced mobility, a longer pipe will lead to more pronounced transmission bias.
This further shows that the inlet capillary should be kept as short as possible, not only to improve absolute quantity of ions transmitted, and thus the sensitivity of the equipment, but also when the relative intensity of the species is considered.

\FloatBarrier

\section{Conclusions}

In our investigation of the ion transport behaviour in heated capillaries of atmospheric pressure interfaces, we limited our study to the cases where the gas flow is fully laminar. 
This condition was found to be the crucial facilitating factor for reaching high transmission ratio.
We further limited the scope to a simple cylindrical geometry. 
The effect of a funnel shaped inlet is modelled by a narrow initial charge distribution  as it  collimates the ion cloud. By this we reproduce the observed unit transmission for not too high ion currents. The classical setup is in contrast modeled by a broad distribution.  

Hence the studied lengths and diameters are not exhaustive, yet representative of capillaries typically used in mass spectrometry equipment and we can identify general trends helpful for instrumentation design.
The maximal transmitted ion current strongly decreases with the capillary length.
Because the gas mass flow decreases to a lesser extent, as defined by the chocking condition, short capillaries are favorable in terms of ions per gas-load ratio.

Certainly, other aspects influence the choice in capillary length, such as reliable droplet evaporation, which is usually supported by some form of heating.
The simulation presented here includes heating in a detailed description of the gas flow, allowing to study its effect on the gas flow and and on the transported ion.
The major finding is that the effect of the heating dramatically depends on the capillary diameter.
Heating leads to dramatically different flows within the 
range of values considered. 
An effectively heated, high viscosity boundary layer contributes little to the overall transport, whereas a cold, low viscosity central gas flow moves most of the ions. 
With increasing diameter this effect becomes more pronounced. 
Thus, for larger diameter capillaries, most transmitted ions are surrounded by cold gas, considerably cooler than atmospheric conditions, due to gas expansion, whereas in the case of small 
diameters, the entire section of the flow is effectively  heated.
Thus ions created by ESI and transported with wide or narrow capillaries might vary due to different thermal stress, which could be crucial for native mass spectrometry.~\cite{Hopper_Nm10(2013)},~\cite{Wolke1131}

Beyond heating, the transmission rate is influenced by  mobility and space charge, both concepts included in the model presented.
Generally, higher mobility yields higher losses.
This basic rationale introduces a bias in the transmission of ions when a mix is considered, in particular when different species compete for transmission of a limited space charge.
As a consequence, high mobility ions surround the low mobility ones and will be lost at the wall at a higher rate. 
This effect is enhanced, in the wide capillary, by the temperature boundary layer close to the wall 
in which the effective mobility is further increased.
This effect has been experimentally highlighted,~\cite{page2009biases} and is reproduced in the presented simulations. This transmission bias, increased in longer capillaries, modifies the proportions that are initially present in the tested sample. It should thus be considered in the applications where relative quantities of species in a mix have to be quantified precisely.

The presented simulation framework offers access to thermodynamical conditions ions are exposed to during 
their transfer into vacuum, including temperature and pressure, but also space charge, collision number and intensity and retention time can be extracted.

The ion transport in interfaces, however, contains further highly relevant aspects, which need to be considered in the future, requiring an even more complex simulation.
Firstly, ion generation from droplets is a crucial step, depending strongly on gas flow and thermodynamical properties.
Technically, the implementation of models predicting physico-chemical transformation of the ions based on those conditions should be easy. 
The knowledge of such models remains however challenging.
Secondly, in order to also consider gas flow beyond the laminar regime, the simulation of a complete three-dimensional flow is needed, as only this allows the description of a turbulent flow and/or the transition between a laminar and a turbulent regime.

---

\section{Supplementary Material} 
\label{resterampe}
 
\subsection{Transmission dependence on creation position} 
\begin{figure}
	\begin{center}
		\includegraphics[width=0.75\linewidth,clip=true]{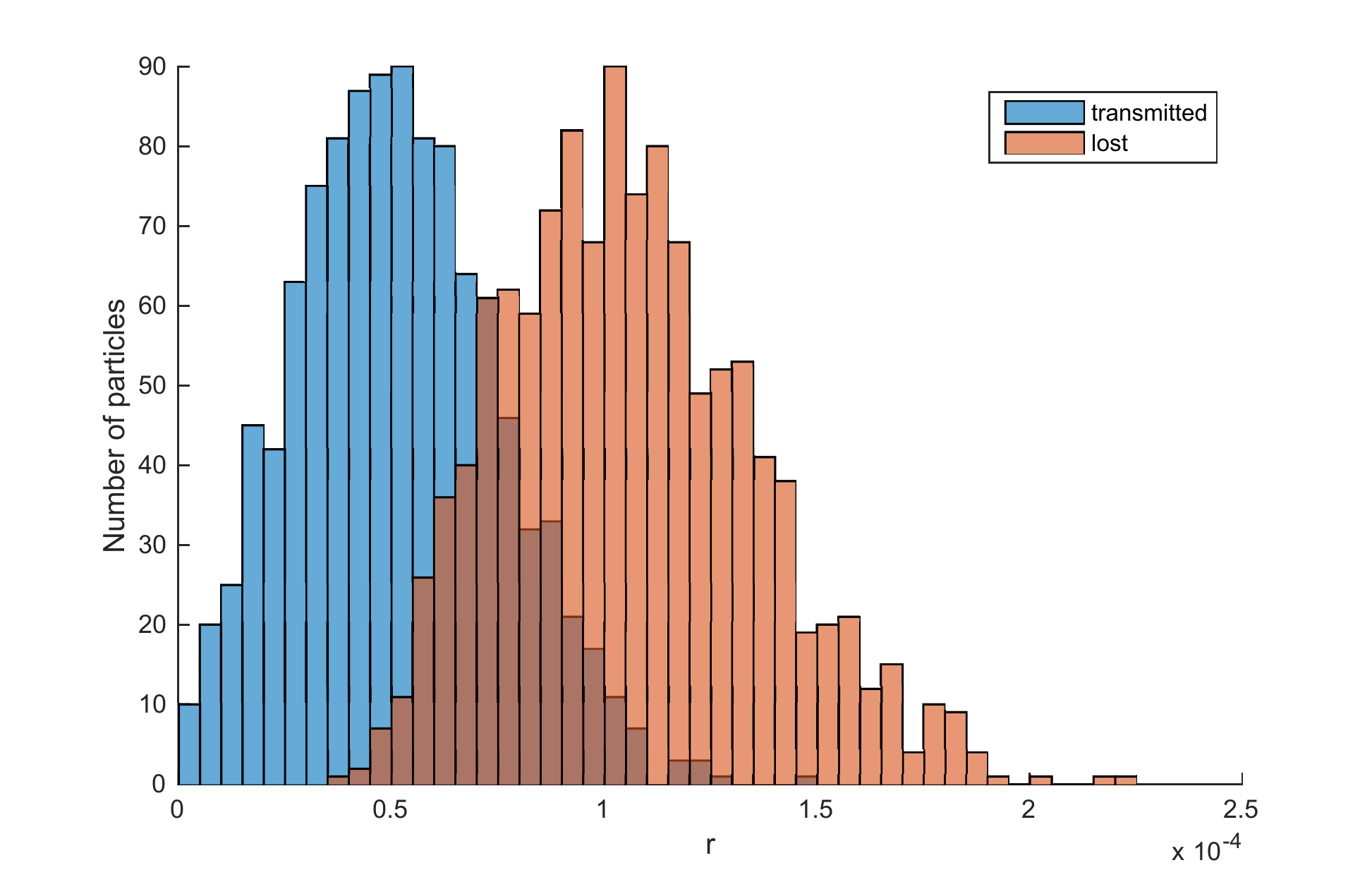}
	\end{center}
	\caption{Repartition of the particles at the generation point, depending on their status: transmitted 
		or lost.
		}
	\label{fig:SMat_story2}
\end{figure}

Looking at the transport from another perspective, further interesting metrics can be obtained. One 
specifically focuses on the initial radial position of the ions. This location corresponds to the 
generation position in the simulation or to the position where the ions enter the capillary in the 
experiments. One follows the particles generated at a given time step, considering the particles close 
enough to the capillary entrance. Those ions are then followed during their flight and ordered as 
transmitted or lost, depending on the last position stored. Figure~\ref{fig:SMat_story2} presents 
the repartition of this initial position depending on those two categories. It appears that the 
three zones can be basically considered. The first zone, close to the symmetry axis of the 
capillary, contains only transmitted ions. The second zone encompasses the largest radial positions 
and contains exclusively lost 
particles. The third domain is the transition between the first two ones, where the proportion of 
transmitted particles decreases whereas the one of lost particles increases when the initial radial 
position increases. This shows that this initial position can be used as a strong indicator of the 
outcome of the journey of the ions in the inlet capillary. This tends to indicate that a good 
concentration of particles around the capillary axis increases enables higher transmissions.

\subsection{Boundary layer}
\begin{figure}
	\begin{center}
		\includegraphics[width=0.75\linewidth,clip=true]{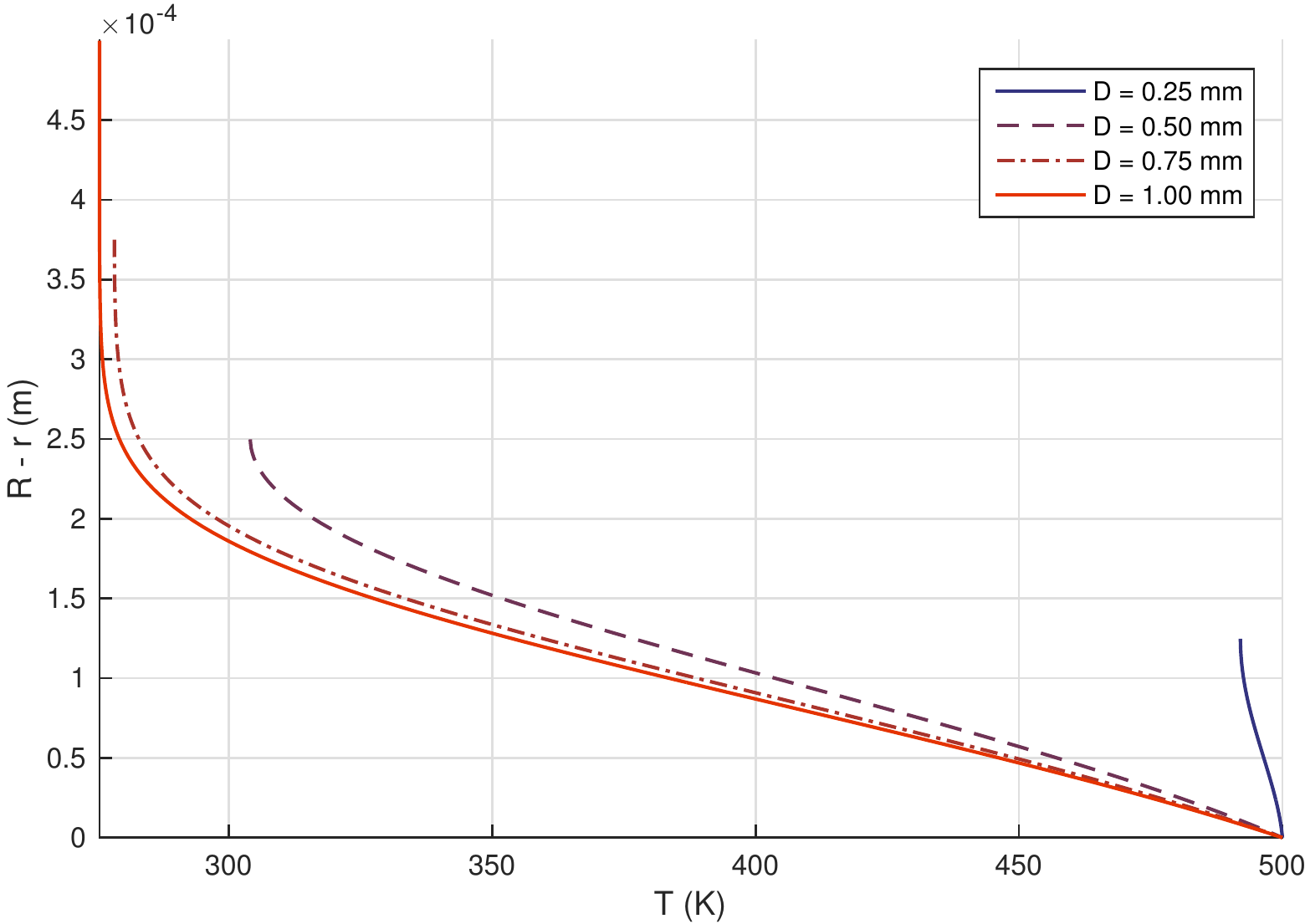}
	\end{center}
	\caption{Influence of the capillary diameter on the boundary layer thickness. All plots are 
		considered at an axial position of $z = \unit{1.5}{\centi\meter}$ for 
		\unit{6.00}{\centi\meter} long capillaries.}
	\label{fig:SMat_TBL}
\end{figure}

A boundary layer in fluid dynamics is defined as an area where the gradient of a given variable is 
specifically high compared to other areas. A typical example is a flow over an infinite plate. Close to 
the plate, the velocity magnitude is zero and far from the plate it reaches a given, non-zero value 
$v_\infty$. Thus, there is an area, close to the plate, where the velocity magnitude varies between 0 
and $v_\infty$. This is the velocity boundary layer. It can be quantitatively defined as the area where 
the velocity magnitude is smaller than $0.99\,v_\infty.$ A similar definition can be used for the 
temperature boundary layer, where the gas temperature reaches a given temperature $T_\textrm{plate}$ 
close to the plate that is different from the temperature far from the plate $T_\infty$.\\
Here we show an exemplary representation of this temperature boundary layer for several 
capillaries, especially focusing on different diameters. In figure~\ref{fig:SMat_TBL}, one can 
see that for the larger diameters, the thickness of the boundary layer is almost identical. This 
thickness being larger than the smallest diameter represented, one can also see that in this case, the 
effects of the heated walls are transmitted up to the center of the domain, and no proper boundary layer 
can be defined.

\subsection{Comparison between radial and axial velocity components}
\begin{figure}
	\begin{center}
		\includegraphics[width=0.75\linewidth,clip=true]{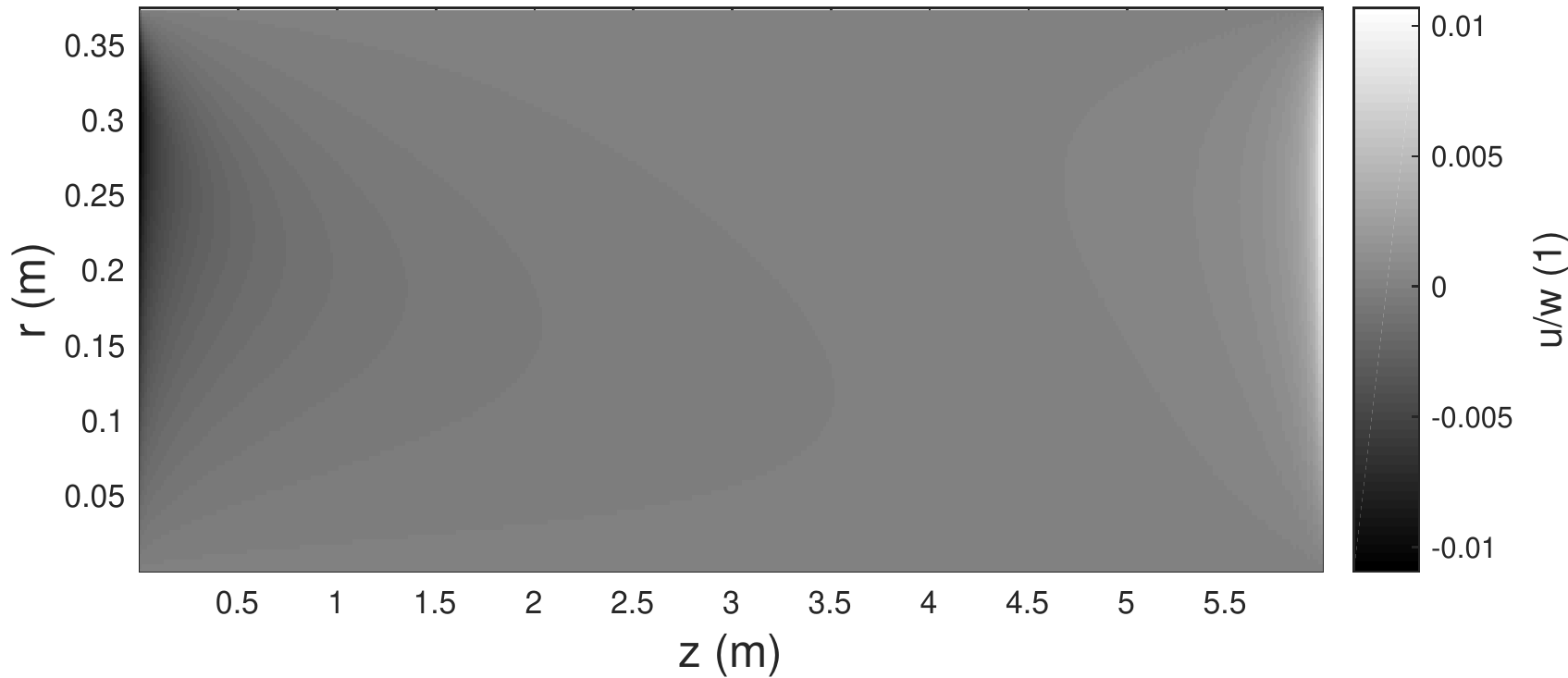}
	\end{center}
	\caption{Ratio between radial and axial velocity components for 
		a \unit{6.00}{\centi\meter} long and \unit{1.00}{\milli\meter} wide capillary 
		(same conditions as for figures~\ref{fig:temperature}-c and~\ref{fig:velocity}-b). Only half of the domain is 
		represented, the other being symmetric.}
	\label{fig:SMat_velRatio}
\end{figure}

We have claimed that the radial velocity component is much smaller than the axial one. To quantify this, 
figure~\ref{fig:SMat_velRatio} represents the ratio between these two components for the same example 
as the one considered in figures~\ref{fig:temperature}-c and~\ref{fig:velocity}-b. One can see that on the entire capillary domain (excluding the walls, as the velocity is set to zero because of the non-slip boundary 
conditions), the radial component of the flow is in the maximal case in the order of 1\% of the axial one.

\subsection{Influence of the pressure set in the outlet section}
\begin{figure}
	\begin{center}
		\includegraphics[width=0.75\linewidth,clip=true]{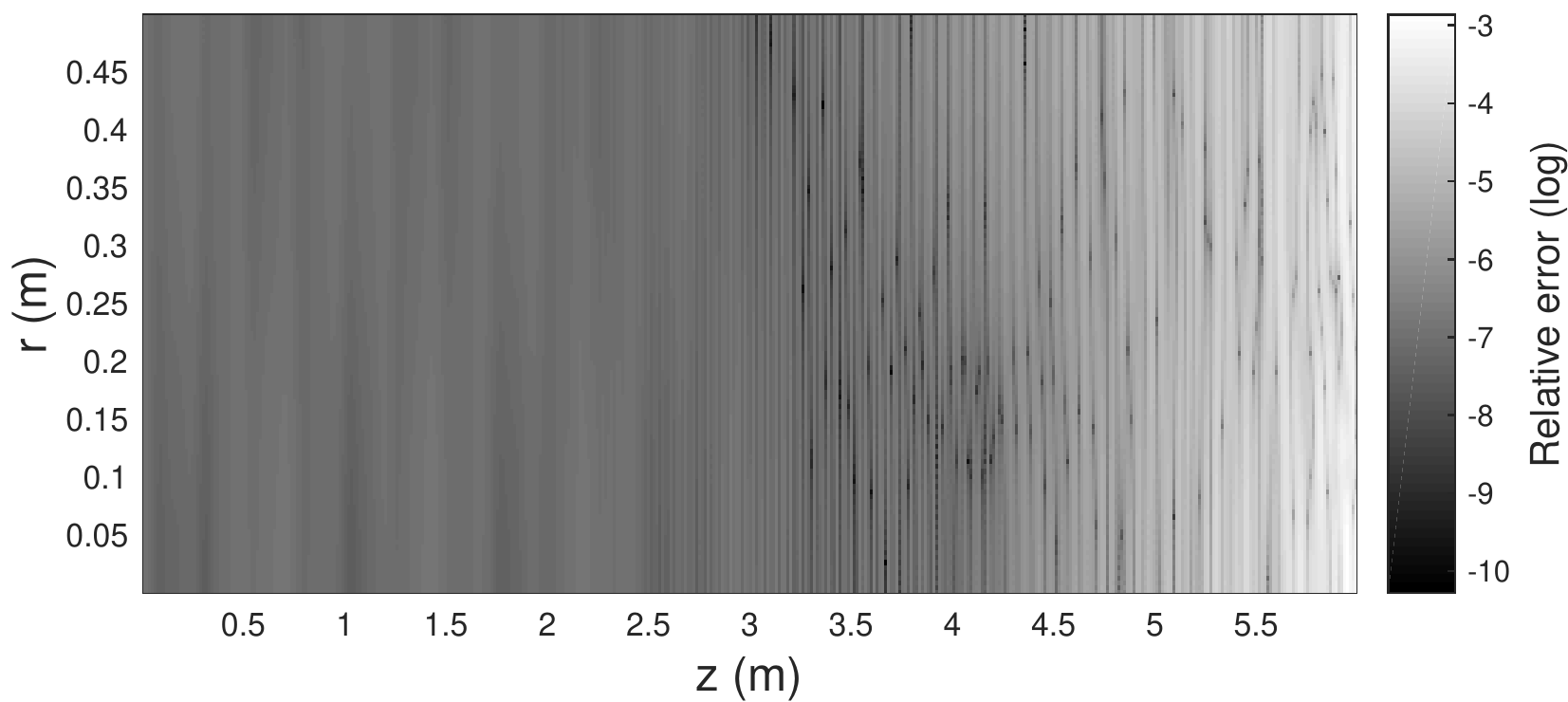}
	\end{center}
	\caption{Relative error in the pressure field for different pressure enforced in the sponge area for 
		\unit{6.00}{\centi\meter} long and \unit{1.00}{\milli\meter} wide capillaries. The pressure 
		enforced in the sponge area is \unit{$1.01325\e{4}$}{\pascal} resp. 
		\unit{$5.06625\e{3}$}{\pascal} (\emph{i.e.} $10~\%$ resp. $5~\%$ of the atmospheric pressure). 
		Only half of the domain is represented, the other being symmetric.}
	\label{fig:SMat_spongeError}
\end{figure}

The pressure difference between both ends of the capillary is simulated through the boundary conditions
applied to the system as described in~\ref{sec:method_flow}. 
The influence of the first vacuum chamber of the machine is modeled via a \emph{sponge} term. It 
modifies the right hand side of the equations, in order to force the pressure field to reach a given value, which has to be chosen for each simulation.
The results should be independent on this parameter, as long as it is lower than the value reached at the 
proper outlet when choking occurs.
This has been tested with different simulations of the same system, where only the enforced pressure in 
the sponge area has been modified. Two cases are presented here, where the pressure is respectively 
\unit{1\cdot\power{10}{4}}{\pascal} and \unit{5\cdot\power{10}{3}}{\pascal}, \emph{i.e.} in both cases 
lower than the choking pressure of about \unit{3\cdot\power{10}{4}}{\pascal}. The maximal relative error 
remains very low in all computed fields: 0.14~\% for the pressure field, 0.3~\% for the density field 
and 0.5~\% for the velocity magnitude. This clearly shows that the maximal errors are encountered close 
to the capillary outlet. This is not surprising as the difference between the simulations is indeed 
located close to this area. This also shows that a large part of the simulation is hardly influenced and 
that even the most concerned areas are only slightly influenced.
The error distribution for the pressure field is presented in Fig.~\ref{fig:SMat_spongeError}.

\subsection{Estimate of the Maximal Ion Current}  
Here we estimate the maximal current passing through a capillary. 
A similar estimate is derived in \cite{skoblin2017numerical}. 
We assume that the ions are introduced as a narrow distribution at start. 
The results hold for the flat distribution, since the outer ions are just lost, and do not influence the electric field close to the axis. This is consistent with the simulations, where the maximal current for both distributions is the same. It might however be difficult to have such high in-currents to reach this saturation for flat distributions in experiment.     

This ion cloud expands, and we assume that the field can be calculated for every cross section independently.

The electric field is 
\begin{equation}
    \pi r \varepsilon_0 E = Q = I /c \,,
\end{equation}
where the electric field $E$, radius of the ion cloud $r$, the vacuum permittivity $\varepsilon_0$, charge per cross section $Q$, ion current $I$ and the mean (axial) flow velocity $u_{\mathrm{av}}$ are introduced. 
The velocity of expansion is given by the field 
\begin{equation}
u_{\mathrm{drift}} = K E  = \frac{K I}{\pi \varepsilon_0 u_{\mathrm{av}} r}\,,
\end{equation}
where $K$ is the mobility of the considered species. $u_{\mathrm{drift}}$ is in turn the change of the radius in time $ dr/dt = u_{\mathrm{drift}}$
\begin{equation}
\frac{dr}{dt} = K E  = \frac{K I}{\pi \varepsilon_0 u_{\mathrm{av}} r}\,.
\end{equation}
Integrating this for a time $T$ and assuming a zero size of the initial cloud we obtain 
\begin{equation}
    r^2 = \frac{2 K I}{\pi \varepsilon_0 u_{\mathrm{av}}} T\,.
\end{equation}
We set $T$ the transmission time, given by $T = L/ u_\mathrm{av}$ and identify $r$ with the radius of the capillary $R$
\begin{equation}
    R^2 = \frac{2 K I L}{\pi \varepsilon_0 u_{\mathrm{av}}^2}\,.
\end{equation}
We find in agreement with the simulations:
\begin{enumerate}
    \item the maximal current is proportional to the cross section ($A\sim R^2$)
    \item if the gas flow velocity does not change, the maximal current is inversely proportional to the length
    \item a reduced mean flow velocity reduces the ion current 
\end{enumerate}
 More precisely, by evaluating the lower boundary of the integration, we obtain: 
 \begin{equation}
    R^2 -r_0^2 = \frac{2 K I L}{\pi \varepsilon_0 u_{\mathrm{av}}^2}\,.
\end{equation}
$r_0 $ is the extension of the initial ion cloud to have unity transmission. It becomes zero for sufficiently large $I$ showing the space charge cut off. The loss rate for a flat distribution is given by
 \begin{equation}
\delta = \frac{I_{\mathrm{trans}}}{I_{\mathrm{in}}} 
= 
\frac{r_0^2}{R^2} 
= 
1 - \frac{2 K I_\mathrm{trans} L}{\pi \varepsilon_0 u_{\mathrm{av}}^2 R^2}\,.
\end{equation}
It can be rewritten in $I_{\mathrm{in}}$ as 
 \begin{equation}
\delta 
= 
1 / \left( 1+  \frac{2 K I_\mathrm{in} L}{\pi \varepsilon_0 u_{\mathrm{av}}^2 R^2}\right)\,.
\end{equation}
Note that the average velocity for the transmission rate  and the velocity for the gas mass flow are not expected to match exactly. The average velocity for the ions  incorporates the flow along the ion trajectories, i.e. the full capillary length. The average gas velocity for the mass flow can be calculated for one cross section, e.g. at the inlet where the density is near constant. 
To match the data we need a smaller $u_{\mathrm{av}}$ for the ion transport than for the  
mass flow by about a factor of two.

\FloatBarrier

\bigskip
This work is supported by the \emph{Deutsche Forschungsgemeinschaft (DFG)} under  \emph{RE-3774/1-1 }

\bibliography{local}{}
\bibliographystyle{unsrt}

\end{document}